\documentclass[aps,prd,showpacs,twocolumn,amssymb]{revtex4}

\usepackage{graphicx}
\usepackage{dcolumn}
\usepackage{bm}
\usepackage{amsmath}
\usepackage{amsfonts}

\begin{document}
\title{Magnetoelectric birefringence revisited}
\author{V. A. De Lorenci}
 \email{delorenci@unifei.edu.br}
\affiliation{Institute of Science, Federal University of Itajub\'a, 
37500-903 Itajub\'a, M. G., Brazil,}
\affiliation{PH Department, TH Unit, CERN, 1211 Geneva 23, Switzerland}
\author{G. P. Goulart}
 \email{gpgoulart@unifei.edu.br}
\affiliation{Institute of Science, Federal University of Itajub\'a, 
37500-903 Itajub\'a, M. G., Brazil.}

\date{\today}

\begin{abstract}
Electromagnetic wave propagation inside isotropic material media 
characterized by dielectric coefficients $\varepsilon_{\mu\nu}(E,B)$ 
and $\mu_{\mu\nu}(E,B)$ is examined. The regime of the eikonal
approximation is considered. The Hadamard 
method of field disturbances is used and the dispersion relations 
are obtained by solving the Fresnel equation. Some applications of the 
formalism are presented. Particularly, birefringence phenomena induced
by applied external fields are derived and discussed. It is shown 
that magnetoelectric birefringence effect can occur even without 
the presence of Kerr and Cotton-Mouton effects, provided the physical
system satisfies certain conditions.
\end{abstract}

\maketitle


%
\section{Introduction}
\label{I}
%

Inside material media the Maxwell equations must be supplemented with constitutive 
relations between the applied external fields and its induced excitations. 
Such relations are generally nonlinear (linear constitutive relations have also 
been considered in the literature \cite{Rubilar}) and depend on the physical 
properties of each considered medium under the action of external fields.  
A remarkable consequence coming from the nonlinearity of the
field equations is the artificially induced birefringence phenomenon: waves with 
different polarization propagate with different velocities due to the presence 
of applied external fields \cite{Souza,Klippert}. 
In fact, the effect of applying an external field is just to induce an artificial
optical axes, which in general disappears as soon as the external field is turned off.
Hence, the initially isotropic medium becomes anisotropic under the action
of external fields.
In the context of crystalline systems the optical axes can be present in the medium
and birefringence occurs naturally \cite{Landau,Born}. In this case, by applying an 
external field both effects (natural and induced birefringences) interfere,
leading to a nontrivial mechanism of birefringence control by means of external 
electromagnetic fields \cite{teodoro2004}. 

In the regime of intense electromagnetic fields the Maxwell linear theory is no longer
applicable and new effects, including birefringence, emerge from the quantum regime 
\cite{Schwinger}. The analysis of light propagation shows that there is a 
non-null probability of photon splitting 
under a strong external electromagnetic field \cite{Birula,Adler}.
Investigations on light propagation in the context of nonlinear
Lagrangian for electrodynamics can be found in  
\cite{Dittrich,DeLorenci,Novello,Gibbons,Refere1}.

Birefringence induced by external electromagnetic fields has been long ago
reported in the literature 
\cite{kerr1875a,kerr1875b,cotton1905a,cotton1905b,cotton1907a,cotton1907b}. 
Distinct proposals for its theoretical description were presented in 
\cite{jones1948,baranova1977,graham1984,ross1989}, where some new aspects 
of this phenomenon were predicted. Among them, the existence of the so called
magnetoelectric birefringence was discussed, which is a kind of birefringence linear
in the product of the electric and magnetic fields. The experimental
observation of the magnetoelectric birefringence was recently reported
in the literature \cite{roth2000,roth2002}.
   
Birefringence is nowadays widely used in the technology of optical
devices as well as a technique for investigating
properties of several physical systems, including astrophysical 
phenomena \cite{prl3}.

In this work, birefringence phenomena are investigated in the context of homogeneous
dielectric media at rest with the dielectric coefficients 
$\varepsilon^\mu{}_\nu = \varepsilon^\mu{}_\nu ({E},{B})$ and
$\mu^\mu{}_\nu = \mu^\mu{}_\nu ({E},{B})$ in the limit of geometrical optics. 
The analysis is restricted to local electrodynamics, where dispersive 
effects are neglected. Only monochromatic waves are considered, thus 
avoiding ambiguities with the velocity of the wave. 

In Section \ref{II} the eigenvalue equation associated with the light
propagation in general non-dispersive material media is presented. 
In Section \ref{III} the dispersion relations are obtained for isotropic
media described by $\varepsilon^\mu{}_\nu = \varepsilon^\mu{}_\nu ({E},{B})$ and
$\mu^\mu{}_\nu = \mu^\mu{}_\nu ({E},{B})$. Applications for birefringence 
phenomena are investigated in Section \ref{IV}, where some particular cases 
are obtained from the formalism. The consequences of the anisotropy 
present in the propagation of the extraordinary ray are discussed.
Some final remarks are presented in the conclusion section. 

A Minkowskian spacetime employing a Cartesian coordinate system is used throughout
this work. The background metric is denoted by $\eta_{\mu\nu} = {\rm diag}(+1,-1,-1,-1)$. 
All quantities are refereed as measured by the geodetic observer $V^\mu=c\delta^\mu_0$,
where $\delta^{\mu}_{\nu}$ denotes the Kronecker tensor.
For any quantity $X^\mu = (0,\vec{X})$ we define its modulus as $X \doteq (-X^{\mu}X_{\mu})^{1/2}$
and the associated unit vector as $\hat{X}\doteq\vec{X}/X$. In the same way we use the
notation $\hat{X}^{\mu}\doteq X^{\mu}/X$. For any two quantities $X^\mu$ and $Y^\mu$ we denote
its scalar product $X^\mu Y_\mu$ as $(XY)$. As an example, the electric field 
is represented by $E^\mu=(0,\,\vec{E})$, whose modulus is $E=(-E^\alpha E_\alpha)^{1/2}$. 
Similarly, the magnetic field is $B^\mu=(0,\,\vec{B})$. Its scalar product is given
by $(EB) = E^\mu B_\mu = -\vec{E}\cdot\vec{B}$.  The units are such that $c=1$.


\section{Eigenvalue equation}
\label{II}

The electrodynamics in a continuum medium at rest is completely determined by the Maxwell
equations \cite{note1}
\begin{eqnarray}
V^{\mu}D^{\alpha}{}_{,\mu}+\eta^{\alpha\beta\gamma\delta}
V_{\gamma}H_{\delta,\beta}&=&0,
\label{1}
\\
V^{\mu}B^{\alpha}{}_{,\mu}-\eta^{\alpha\beta\gamma\delta}
V_{\gamma}E_{\delta,\beta}&=&0,
\label{2}
\end{eqnarray}
together with the constitutive relations 
$D^{\alpha} = \varepsilon^{\alpha}{}_{\beta}E^{\beta}$
and 
$H^{\alpha} = \mu^{\alpha}{}_{\beta}B^{\beta}$.
The coefficients $\varepsilon^{\alpha}{}_{\beta}=\varepsilon^{\alpha}{}_{\beta}
(E^\mu,\,B^\mu)$ and  $\mu^{\alpha}{}_{\beta}=\mu^{\alpha}{}_{\beta}
(E^\mu,\,B^\mu)$ represent the dielectric tensors. They are usually denoted
as permittivity and permeability tensors, respectively. All the information 
about the dielectric properties of the medium can be encompassed in it. 

The propagation of the electromagnetic waves in the eikonal 
approximation of electrodynamics \cite{Landau} is obtained by making use of the method of 
field disturbances \cite{Plebanski,Hadamard}. With the notation introduced 
before \cite{Souza} we set   
$\left[E^\mu{}_{,\nu}\right]_{\Sigma} = e^{\mu}K_{\nu}$ and
$\left[B^\mu{}_{,\nu}\right]_{\Sigma} = b^{\mu}K_{\nu},$
where $e^{\mu}$ and $b^{\mu}$ are related to the derivatives of the electric and magnetic
fields on the surface of discontinuity $\Sigma$, and they are associated with the 
polarization of the propagating waves. The quantity $K_{\lambda}=\partial \Sigma / \partial x^\lambda$ 
is the wave 4-vector normal to $\Sigma$. 
Applying these boundary conditions to the field equations (\ref{1}) and 
(\ref{2}), we obtain the eigenvalue equation \cite{teodoro2004}
\begin{equation}
Z^\alpha{}_\tau e^\tau=0,
\label{5}
\end{equation}
where the Fresnel tensor $Z^{\alpha}{}_{\tau}$ is given by
%
\begin{eqnarray}
Z^\alpha{}_\tau &\doteq& \,C^{\alpha}{}_\tau 
+ \frac{1}{\omega}\frac{\partial \varepsilon^{\alpha}{}_{\beta}}
{\partial B^{\chi}}\eta^{\chi\omega\gamma}{}_{\tau}E^{\beta}V_\gamma q_\omega
\nonumber
\\
&& +\frac{1}{\omega}\frac{\partial\mu^{\delta}{}_{\lambda}}{\partial E^\tau}
\eta^{\alpha\beta\gamma}{}_{\delta} B^{\lambda} V_\gamma q_\beta 
+ \frac{1}{\omega^2}\left(q^2 H_\omega{}^\alpha I^\omega{}_\tau 
\right.
\nonumber
\\
&& \left. - q^2 H_\chi{}^\chi I^\alpha{}_\tau + H_\tau{}^\beta q_\beta q^\alpha 
- H_\chi{}^\delta q^\chi q_\delta h^\alpha{}_\tau \right),
\label{6}
\end{eqnarray}
with $\omega \doteq K^\alpha V_\alpha$ representing the angular frequency of the
electromagnetic wave. Additionally, the following definitions are introduced:
\begin{eqnarray}
h^\alpha{}_\tau &\doteq& \delta^\alpha{}_\tau - V^\alpha V_\tau,
\label{11}
\\
q^\alpha &\doteq& h^\alpha{}_\tau K^\tau = K^\alpha - \omega V^\alpha, 
\label{9}
\\
q^2 &=& -q^{\alpha}q_{\alpha} = \omega^2 - K^2,
\label{9b}
\\
C^{\alpha}{}_\tau &\doteq& \varepsilon^{\alpha}{}_{\tau} +
\frac{\partial \varepsilon^{\alpha}{}_{\beta}}{\partial E^{\tau}}E^\beta,
\label{*}
\\
H^{\alpha}{}_\tau &\doteq& \mu^{\alpha}{}_{\tau} + 
\frac{\partial \mu^{\alpha}{}_{\beta}}{\partial B^{\tau}}B^{\beta},
\label{8}
\\
I^\alpha{}_\tau &\doteq& h^\alpha{}_\tau + \frac{q^\alpha q_\tau}{q^2}.
\label{7}
\end{eqnarray}

The general solution for the wave propagation can be derived from 
the eigen-value problem stated by the generalized Fresnel equation 
[Eq.\ (\ref{5})], and is formally given by 
$
\det|Z^\alpha{}_\beta| = 0.
\label{1*}
$
There are different ways to solve this problem. In the next section an
expansion of the polarization vector $e^\mu$ in a suitable basis of linearly
independent vectors will be considered in order to obtain solutions from 
the eigen-value equation.

\section{The dispersion relations}
\label{III}

We shall examine the wave propagation issue in material media with dielectric coefficients 
depending on the intensity of the external fields. For this case we set:
\begin{eqnarray}
\varepsilon^{\alpha}{}_{\tau} &=& \varepsilon (E,B) h^{\alpha}{}_{\tau},
\label{14}\\
\mu^{\alpha}{}_{\tau} &=& \mu^{-1} (E,B) h^{\alpha}{}_{\tau}.
\label{15}
\end{eqnarray}
Since these coefficients do not depend on the directions of the external fields, 
the results are applicable to the study of wave propagation in isotropic liquid media. 
The application of the results to crystalline structures can be done by considering
the explicit dependence  of $E^\mu$ or $B^\mu$ vector fields on the dielectric coefficients.
These cases were partially described in \cite{Souza,teodoro2004}. 

From Eqs. (\ref{14}) and (\ref{15}), we obtain:
\begin{eqnarray}
\frac{\partial \varepsilon^\alpha{}_{\beta}} {\partial E^\tau} &=& 
- {}\varepsilon'{} h^{\alpha}{}_{\beta} E_{\tau}, 
\qquad\;\;\, 
\frac{\partial \varepsilon^\alpha{}_{\beta}} {\partial B^\tau} = 
- {}\dot{\varepsilon}{} h^{\alpha}{}_{\beta} B_{\tau}, 
\label{17a}\\
\frac{\partial \mu^\alpha{}_{\beta}} {\partial E^\tau} &=& 
+ {}\frac{\mu'}{\mu^2}{} h^{\alpha}{}_{\beta} E_{\tau}, 
\qquad 
\frac{\partial \mu^\alpha{}_{\beta}} {\partial B^\tau} = 
+ {}\frac{\dot{\mu}}{\mu^2}{} h^{\alpha}{}_{\beta} B_{\tau},  
\label{17}
\end{eqnarray}
where we have defined 
$X' \doteq (1/E)\partial X/\partial E$ and
$\dot{X} \doteq (1/B)\partial X/\partial B$,
for any quantity $X$.

By introducing the last results in $Z^\alpha{}_{\beta}$, from Eq. (\ref{6}), it yields:
%
\begin{eqnarray}
Z^\alpha\,_\tau &=& 
\left[\varepsilon -\frac{q^2}{\mu\omega^2} - \frac{\dot{\mu}q^2}{\mu^2\omega^2}
(BIB)\right] h^{\alpha}{}_{\tau} 
- \varepsilon' E^\alpha E_\tau  
\nonumber
\\
&&
+ \frac{1}{\omega^2}\,\left(\frac{\dot{\mu}B^2}{\mu^2} 
- \frac{1}{\mu}\right) q^\alpha q_\tau 
+ \frac{\dot{\mu}q^2}{\mu^2\omega^2} B^\alpha B_\tau 
\nonumber
\\
&&
+ \frac{\dot{\mu}}{\mu^2\omega^2} (qB) B^\alpha q_\tau 
+ \frac{\dot{\mu}}{\mu^2\omega^2} (qB) q^\alpha B_\tau 
\nonumber
\\
&&
- \left[\frac{\dot{\varepsilon}}{\omega} \eta^{\beta\gamma\delta}{}_{\tau}E^\alpha  
+ \frac{\mu'}{\omega\mu^2}\eta^{\beta\gamma\delta\alpha}E_\tau\right]  q_\beta V_\gamma B_\delta 
\label{23}
\end{eqnarray}
%
where the following additional notation was introduced: 
\begin{eqnarray}
(XIY) &\doteq& X_\mu I^{\mu\nu} Y_\nu = (XY) + \frac{(qX)(qY)}{q^2},
\label{XIY}
\end{eqnarray}
for any quantities $X$ and $Y$.

In the present form, it is clear that $Z^\alpha{}_\tau$ is a 3-dimensional object. 
Particularly we note the $Z^\alpha{}_\tau V^\tau = 0$. In order to find solutions of the 
eigen-value equation, Eq. (\ref{5}) with $Z^\alpha{}_\tau$ given by Eq. (\ref{23}), we shall
consider the expansion of the polarization vector $e^\tau$ in a convenient basis of 
the 3-dimensional space as 
\begin{eqnarray}
e^\tau = a \,E^\tau + b\, B^\tau + c\, q^\tau.
\label{24}
\end{eqnarray}
In this case, from Eqs.(\ref{5}) and (\ref{23}), we obtain
\begin{widetext}
\begin{eqnarray}
&&\left\{a \left[
\varepsilon -\frac{q^2}{\mu\omega^2} - \frac{\dot{\mu}q^2}{\mu^2\omega^2}(BIB)
+ \varepsilon' E^2 - \frac{\dot{\varepsilon}}{\omega} 
\eta^{\beta\gamma\delta\tau} q_\beta V_\gamma B_\delta E_\tau 
+ \frac{\alpha\mu'E^2}{\omega\mu^2}\right] + b \left[-\varepsilon'(EB) 
- \frac{\alpha\mu'(EB)}{\omega\mu^2}\right]\right. 
\nonumber\\
&+&\left. c \left[-\varepsilon'(qE)
- \frac{\alpha\mu'(qE)}{\omega\mu^2}\right] \right\} E^\alpha  
+ \left\{a\left[\frac{\dot{\mu}q^2 (EB)}{\omega^2\mu^2} 
+ \frac{\dot{\mu}(qB)(qE)}{\omega^2\mu^2} + \frac{\beta\mu'E^2}{\omega\mu^2}\right] 
+ b \left[\varepsilon -\frac{q^2}{\mu\omega^2} - \frac{\beta\mu'(EB)}{\omega\mu^2}\right]\right. 
\nonumber\\
&+& \left. c \left[-\frac{\beta\mu'(qE)}{\omega\mu^2}\right]\right\} B^\alpha 
+ \left\{a\left[\frac{1}{\omega^2}\left(\frac{\dot{\mu}B^2}{\mu^2} 
- \frac{1}{\mu}\right)(qE) + \frac{\dot{\mu}(qB)(EB)}{\omega^2\mu^2} 
+ \frac{\gamma\mu' E^2}{\omega\mu^2}\right]\right. 
\nonumber\\
&+& \left. b \left[\frac{1}{\omega^2}\left(\frac{\dot{\mu}B^2}{\mu^2} - \frac{1}{\mu}\right)(qB) 
- \frac{\dot{\mu}B^2(qB)}{\omega^2\mu^2} - \frac{\gamma\mu' (EB)}{\omega\mu^2}\right] 
+ c\left[\varepsilon - \frac{\gamma\mu' (qE)}{\omega\mu^2}\right]\right\} q^\alpha = 0,
\label{46}
\end{eqnarray}
\end{widetext}
where $\alpha$, $\beta$ and $\gamma$ are given by
\begin{eqnarray}
\alpha &=& \dfrac{q(BIB)}{\sqrt{(EIE)(BIB) - (EIB)^2}},
\label{42}
\\
\beta &=& \dfrac{-q(EIB)}{\sqrt{(EIE)(BIB) - (EIB)^2}},
\label{43}
\\
\gamma &=& \dfrac{(qE)(BIB) - (qB)(EIB)}{q\sqrt{(EIE)(BIB) - (EIB)^2}}.
\label{44}
\end{eqnarray}
Since $\{E^\alpha,B^\alpha,q^\alpha\}$ are taken to be linearly independent vectors, 
we obtain the solution to the above system by setting each of its coefficients to zero. 
With a convenient notation, it yields the following system of algebraic equations:
\begin{eqnarray}
a \left\{A_1\right\} - b \left\{B_1\right\} - c \left\{C_1\right\} &=& 0,
\label{a1}\\
a \left\{A_2\right\} + b \left\{B_2\right\} - c \left\{C_2\right\} &=& 0,
\label{a2}\\
a \left\{A_3\right\} - b \left\{B_3\right\} + c \left\{C_3\right\} &=& 0,
\label{a3}
\end{eqnarray}
where
\begin{eqnarray}
A_1 &\doteq& \varepsilon + \left(\varepsilon' + \frac{\alpha\mu'}{\omega\mu^2}\right) E^2 
- \frac{q^2}{\mu\omega^2}\left[1 + \frac{\dot{\mu}}{\mu}\,(BIB)\right] 
\nonumber\\
&&- \frac{\dot{\varepsilon}}{\omega}\,\left[qVBE\right], 
\label{59}
\\
A_2 &\doteq& \frac{\dot{\mu}q^2(EIB)}{\omega^2\mu^2} + \frac{\beta\mu' E^2}{\omega\mu^2},
\label{60}
\\
A_3 &\doteq& \frac{\dot{\mu}}{\omega^2\mu^2}\left[B^2(qE)+(qB)(EB)\right] 
- \frac{(qE)}{\mu\omega^2} 
\nonumber
\\
&&+ \frac{\gamma\mu'E^2}{\omega\mu^2}, 
\label{61}
\end{eqnarray}
\begin{eqnarray}
B_1 &\doteq& (EB)\left(\varepsilon' + \frac{\alpha\mu'}{\omega\mu^2}\right), 
\label{62}
\\
B_2 &\doteq& \varepsilon - \frac{q^2}{\mu\omega^2} - \frac{\beta\mu'(EB)}{\omega\mu^2}, 
\label{63}
\\
B_3 &\doteq& \frac{(qB)}{\mu\omega^2} + \frac{\gamma\mu'(EB)}{\omega\mu^2}, 
\label{64}
\end{eqnarray}
\begin{eqnarray}
C_1 &\doteq& (qE)\left(\varepsilon' + \frac{\alpha\mu'}{\omega\mu^2}\right), 
\label{65}
\\
C_2 &\doteq& \frac{\beta\mu'(qE)}{\omega\mu^2}, 
\label{66}
\\
C_3 &\doteq& \varepsilon - \frac{\gamma\mu'(qE)}{\omega\mu^2},
\label{67}
\end{eqnarray}
with $[qVBE] \doteq \eta^{\alpha\beta\gamma\delta}\,q_\alpha V_\beta B_\gamma E_\delta
= \vec{q}\cdot (\vec{E}\times\vec{B})$.

The above set of algebraic equations [Eqs. (\ref{a1})-(\ref{a3})] 
can be solved immediately, resulting in:
\begin{eqnarray}
&&\left(A_3 B_2 + A_2 B_3\right) C_1 + \left(A_3 B_1 - A_1 B_3\right) C_2  
\nonumber
\\
&&+ \left(A_1 B_2 + A_2 B_1\right) C_3 = 0.
\label{58}
\end{eqnarray}
%
This is the general equation governing the phenomenon of electromagnetic wave 
propagation inside material media described by the dielectric coefficients given by Eqs. (\ref{14}) 
and (\ref{15}). Such equation is usually called as the dispersion relation. In the next
section some electro-magnetic-optic effects will be derived from it and known
results concerning birefringence phenomena will be recovered and discussed.

For several physical configurations, the dispersion relations can be presented 
in the suggestive form $g_\pm^{\mu\nu}K_\mu K_\nu = 0$. The symmetric tensors 
$g_\pm^{\mu\nu}$ represents the optic metrics and, generally, present a solution for
each possible polarization mode inside the medium. 
The integral curves of the vector $K_{\mu}$ are geodesics in the associated effective 
geometry \cite{Klippert}. 

Before closing this section some comments on the applicability of the vector basis introduced in
Eq. (\ref{24}) are in order. In the next section we shall apply the results obtained
here to study some limiting cases in which two of the vectors used in the basis are parallel.
In fact, even for those cases the dispersion relation stated by Eq. (\ref{58}) holds.
It can be understood as follows. Let us consider, for instance, an specific configuration
where the angle $\epsilon$ between the vectors $\hat{E}$ and $\hat{B}$ is small. 
Thus, the products between these vectors can be presented as
$\hat{E}\cdot\hat{B} \simeq 1 -\epsilon^2/2$ and $\|\hat{E}\times\hat{B}\| \simeq \epsilon$.
For $\epsilon \neq 0$ the set of vectors introduced in Eq. (\ref{24}) will 
still be a set of basis vectors. Now, if we take $\epsilon$
to be sufficiently small, such that its contribution in the dispersion relation is not measurable, the obtained results
must be the same as if $\hat{E}$ and $\hat{B}$ would be parallel vectors.
In this way, the results derived in this section can be naturally extended to deal with
these limiting cases. 

In fact, the dispersion relation obtained in this section can also
be obtained without taking in consideration an specific vector basis. Using the
Cayley-Hamilton method (see, for instance,  Ref. \cite{rodrigues}), the eigenvalue 
equation can be solved by means of
$(Z_1)^3-3Z_1Z_2+2Z_3=0$, where $Z_i$ ($i = 1,2,3$) represents the traces of tensor $Z^\mu{}_\nu$.
For simple cases, as occur in the derivation of the Kerr effect, the use of the Cayley-Hamilton
method is convenient. Nevertheless, for more elaborate situations, as it occurs in the
magnetoelectric birefringences, the use of this method leads to laborious calculations.
As simple application of this method the Kerr effect is derived in the appendix.

\section{Applications to birefringence phenomena}
\label{IV}
In this section some specific examples will be derived from the general dispersion
relation presented in Eq. (\ref{58}). First we will recover some well know cases, as
the Kerr electro-optical and the Cotton-Mouton magneto-optical effects,
in which the birefringence phenomena appear as a consequence of the applied external
fields. After we will describe the recently measured \cite{roth2000,roth2002}
magneto-electric birefringence. In particular, the anisotropic behavior of the
extraordinary ray with respect to the direction of propagation will be carefully
examined.

\subsection{Electro and magneto-optical effects}
\label{iva}
Let us now consider the particular case where $\varepsilon = \varepsilon(E)$ and 
$\mu = \mu(B)$. These functional dependence of the dielectric coefficients are,
respectively, the necessary conditions to the existence of the Kerr and the 
Cotton-Mouton birefringence effects in material media, as described by the present 
formalism. 

For this case, it follows from Eq. (\ref{58}) together with definitions stated 
in Eqs. (\ref{59}) - (\ref{67}), that:
\begin{eqnarray}
\Lambda_1 v^4 + \Lambda_2 v^2 + \Lambda_3 = 0,
\label{89}
\end{eqnarray}
where we have defined the phase velocity $v^2 = {\omega^2}/{q^2}$, and
%
\begin{eqnarray}
&\Lambda_1&\!\!\! \doteq \varepsilon^2 (\varepsilon + \varepsilon' E^2),
\label{91}
\\
&\Lambda_2&\!\!\! \doteq \frac{\varepsilon\,\varepsilon'\dot\mu}{\mu^2} 
\left[B^2(\hat q E)^2 
+ (EB)^2 + 2(EB)(\hat q E)(\hat q B)\right] 
\nonumber
\\
&&\!\!\! - \frac{\varepsilon^2\dot\mu(BIB)}{\mu^2} - \frac{2\varepsilon^2 
+ \varepsilon\varepsilon' E^2 + \varepsilon\,\varepsilon'(\hat q E)^2}{\mu}, 
\label{92}
\\
&\Lambda_3&\!\!\! \doteq \frac{\varepsilon'\dot \mu (BIB)(\hat q E)^2 
+ \varepsilon \dot\mu (BIB)}{\mu^3}
+ \frac{\varepsilon + \varepsilon'(\hat q E)^2}{\mu^2}. 
\label{93}
\end{eqnarray}
%
Finally, we obtain the following quadratic equation for the phase velocity:
\begin{eqnarray}
v^2 = \dfrac{-\Lambda_2 \pm \sqrt{\Lambda_2^2 - 4 \Lambda_1 \Lambda_3}}{2 \Lambda_1}. 
\label{94}
\end{eqnarray}
In general, we can obtain two solutions from the above equation, which correspond
to the two possible polarization modes propagating in the medium. 

\subsubsection{Kerr birefringence}
\label{iva1}
By considering $\mu=\mu_c = $ constant in the Eq. (\ref{94}) we obtain the following
solutions:
\begin{eqnarray}
v_o^2 = \frac{1}{\mu_c\varepsilon}, 
\label{71}
\end{eqnarray}
and 
\begin{eqnarray}
v_e^2 = \frac{1}{\mu_c (\varepsilon + \varepsilon' E^2)} \left[1 
+ \frac{E}{\varepsilon} \frac{\partial\, \varepsilon}{\partial E} (\hat{q}\cdot\hat{E})^2\right],
\label{74}
\end{eqnarray}
where the index $o$ stands for the ordinary ray ($o$-ray), which propagates isotropically, and
the index $e$ stands for the extraordinary ray ($e$-ray), which depends on the direction of
wave propagation. Particularly, if the propagation occurs parallel to the external 
electric field, $(\hat{q}\hat{E})=\hat{q}\cdot\hat{E} = 1$, the velocities of both rays 
coincide. The difference 
between them achieve its maximum value when $(\hat{q}\hat{E})=\hat{q}\cdot\hat{E} = 0$. 
In this case Eq. (\ref{74}) results in:
\begin{eqnarray}
v_{e\bot}^2 = \frac{1}{\mu_c (\varepsilon + \varepsilon' E^2)}. 
\label{77}
\end{eqnarray}
By considering the expansion of the permittivity as
\begin{eqnarray}
\varepsilon = \varepsilon_c + \varepsilon_1 \,E^2, 
\label{78}
\end{eqnarray}
with $\varepsilon_c$ and $\varepsilon_1$ constants, we obtain
\begin{eqnarray}
&&v_{e\|}^2 = v_o^2=\frac{1}{\mu_c\,\varepsilon_c (1 + \varepsilon_1\,E^2/\varepsilon_c)}, 
\label{79}\\
\nonumber\\
&&v_{e\bot}^2 = \frac{1}{\mu_c\,\varepsilon_c (1 + 3 \,\varepsilon_1\,E^2/\varepsilon_c)}. 
\label{80}
\end{eqnarray}
It is assumed that $\varepsilon_1 E^2$ corresponds to a small correction compared to the 
background permittivity $\varepsilon_c$. The corresponding refraction indexes 
($n=1/v$) are given by:
\begin{eqnarray}
&&n_{\|} = \frac{1}{v_{e\|}} \simeq  
\sqrt{\varepsilon_c\mu_c} \left(1 + \frac{\varepsilon_1\,E^2}{2\,\varepsilon_c}\right),
\\
&&n_{\bot} = \frac{1}{v_{e\bot}} \simeq 
\sqrt{\varepsilon_c\mu_c} \left(1 + \frac{3\,\varepsilon_1\,E^2}{2\,\varepsilon_c}\right).
\end{eqnarray}
Finally the observable quantity defined as the maximum difference between the refraction indexes
in the system is given, up to second order corrections, by 
\begin{eqnarray}
n_{\bot} - n_{\|} = \sqrt{\varepsilon_c\,\mu_c} \;\frac{\varepsilon_1\,E^2}{\varepsilon_c}.
\label{81}
\end{eqnarray}
This result is known as the Kerr electro-optic effect. For this case, the presence of an 
external magnetic field does not produce any change in the results. 

\subsubsection{Cotton-Mouton birefringence}
Now, taking $\varepsilon = \varepsilon_c =$ constant in Eq. (\ref{94}), we obtain 
the following solutions:
\begin{eqnarray}
v_o^2 = \frac{1}{\varepsilon_c\, \mu} 
\label{71b}
\end{eqnarray}
and 
\begin{eqnarray}
v_e^2 = \frac{1}{\varepsilon_c\, \mu} 
\left[1 + \frac{\dot{\mu}}{\mu}(BIB)\right].
\label{74b}
\end{eqnarray}
As before, the velocity of the $e$-ray depends on the direction of propagation. In the particular 
case where the propagation occurs parallel or anti-parallel to the external 
magnetic field ($\hat{q}\cdot\hat{B} = \pm 1$), 
its velocity reduces to the velocity of the $o$-ray. The difference between the velocities of the ordinary and 
extraordinary rays achieve its maximum value when $\hat{q}\cdot\hat{B} = 0$. In this 
case, from Eq. (\ref{74b}):
\begin{eqnarray}
v_{e\bot}^2 = \frac{1}{\varepsilon_c \mu} 
\left(1 - \frac{\dot{\mu}}{\mu}B^2\right).
\label{77b}
\end{eqnarray}
By considering the expansion of the permeability as
\begin{eqnarray}
\mu =  \mu_c + \varepsilon_2 \,B^2, 
\label{78b}
\end{eqnarray}
with $\mu_c$ and $\varepsilon_2$ constants, we obtain
\begin{eqnarray}
&&v_{e\|}^2 = v_o^2= \frac{1}{\mu_c\,\varepsilon_c} \left(1 - \frac{\varepsilon_2 \,B^2}{\mu_c}\right), 
\label{79b}\\
\nonumber\\
&&v_{e\bot}^2 = \frac{1}{\mu_c\,\varepsilon_c} \left(1 - \frac{3\varepsilon_2 \,B^2}{\mu_c}\right). 
\label{80b}
\end{eqnarray}
By considering $\varepsilon_2 B^2 << \mu_c$, the corresponding refraction indexes are given by
\begin{eqnarray}
&&n_{\|} \simeq \sqrt{\varepsilon_c\,\mu_c} \left(1 + \frac{\varepsilon_2\,B^2}{2 \mu_c}\right),
\\
&&n_{\bot} \simeq \sqrt{\varepsilon_c\,\mu_c} \left(1 + \frac{3\,\varepsilon_2\,B^2}{2\,\mu_c}\right).
\end{eqnarray}
Now the difference between these refraction indexes is given by
\begin{eqnarray}
n_{\bot} - n_{\|} \simeq \sqrt{\varepsilon_c\,\mu_c} \;\frac{\varepsilon_2\,B^2}{\mu_c}. 
\label{81b}
\end{eqnarray}
This result is known as the Cotton-Mouton magneto-optic effect. The presence of an
external electric field is not important for this case. 

These two effects, Kerr and Cotton-Mouton birefringences, are
really symmetric effects. This symmetry is related to the fact that the 
Maxwell equations in the absence of sources are symmetric with respect to 
the duality rotation.

\subsection{Magnetoelectric birefringence}
\label{ivb}

As another application of the present formalism, let us now examine some features of the
so called magnetoelectric birefringence. For simplicity, we shall assume in this
section the particular case where $\varepsilon = \varepsilon(E,B)$ and  $\mu = \mu_c =$ constant.

In this case, from Eq. (\ref{58}) we obtain the following solutions for 
the phase velocities of both the ordinary and the extraordinary rays:
\begin{eqnarray}
v_o = \pm \frac{1}{\sqrt{\mu_c\,\varepsilon}},&& 
\label{129}
\end{eqnarray}
and
\begin{eqnarray}
v_e{}^{\pm} \! =\! \dfrac{\dot{\varepsilon}\left[\hat{q}VBE\right]}{2(\varepsilon 
\!+\! \varepsilon'E^2)} 
\pm \sqrt{\frac{\dot{\varepsilon}^2 \left[\hat{q}VBE\right]^2}{4(\varepsilon
\!+\!\varepsilon'E^2)^2}\! +\! \frac{\varepsilon \!+\! \varepsilon'(\hat{q}E)^2}{\mu\varepsilon
(\varepsilon\!+\!\varepsilon'E^2)}}.
\label{101}
\end{eqnarray}
As one can see, the equation for the velocity of the $e$-ray contains two possible solutions, which are not
obviously symmetric with respect to the direction of propagation. Together with the
$o$-ray we generally have three distinct solutions for the phase velocities. In fact the two 
solutions associated with the $e$-ray indicate that it can present distinct velocities in 
opposite directions. This aspect will be 
further addressed in the analysis of some particular cases. 

Let us consider the following expansion for the permittivity coefficient 
\begin{equation}
\varepsilon = \varepsilon_c + \varepsilon_1 E^2 + \varepsilon_2 B^2 
+ \varepsilon_3 (\vec{E}\cdot \vec{B}).
\label{102}
\end{equation}
For practical cases, the terms in $\varepsilon_i$ (i=1,2,3) represent just
small corrections to the main term $\varepsilon_c$ -- the background permittivity. By 
using this expansion we obtain the following results for the velocities of the above rays,
up to second order terms in $\varepsilon_i$,
\begin{equation}
v_o =\! \pm\! \frac{1}{\sqrt{\varepsilon_c\mu_c}}\left[1\! -\! \frac{\varepsilon_1}{2\varepsilon_c}E^2 
\!-\! \frac{\varepsilon_2}{2\varepsilon_c}B^2 \!-\! \frac{\varepsilon_3}{2\varepsilon_c}EB 
(\hat{E}\cdot\hat{B})\right],
\label{131}
\end{equation}
and
\begin{eqnarray}
v_e{}^{\!\pm} &=& \left[\frac{\varepsilon_2}{\varepsilon_c}EB + \frac{\varepsilon_3}{2\varepsilon_c}
E^2(\hat{E}\cdot\hat{B})\right]\hat{q}\cdot(\hat{E}\times\hat{B})
\nonumber
\\
&&
\pm\frac{1}{\sqrt{\varepsilon_c\mu_c}}\left\{1 + \left[(\hat{q}\cdot\hat{E})^2-\frac{3}{2}\right]
\frac{\varepsilon_1}{\varepsilon_c}E^2 - \frac{\varepsilon_2}{2\varepsilon_c}B^2 
\right.
\nonumber
\\
&&
\left. 
+\frac{\varepsilon_3}{2\varepsilon_c}EB(\hat{E}\cdot\hat{B})\left[(\hat{q}\cdot\hat{E})^2-2\right]
\right\}.
\label{105}
\end{eqnarray}

Now, two particular cases of current experimental interest are analyzed. First we 
shall derive an example of the recently measured magnetoelectric linear birefringence 
\cite{roth2002}. After, the so called Jones effect will be presented. The latter is also a
kind of magnetoelectric birefringence which was predicted long ago \cite{jones1948} and was recently
measured \cite{roth2000}.  

\subsubsection{Some specific cases}
Let us consider the case where the direction of the wave vector $\hat{q}$ is perpendicular to the plane
which contains the electric $\vec{E}$ and magnetic $\vec{B}$ fields. 
Let us work in a Cartesian coordinate
system and set $\hat{E}\cdot\hat{B}=\cos\theta$ and $\hat{q}=\hat{z}$ such that  
$\hat{q} = {(\hat{E}\times\hat{B})}/{\|\hat{E}\times\hat{B}\|}$.
Up to second order terms, the velocity of the $e$-ray, given by Eq. (\ref{105}), is
\begin{equation}
v_e{}^{\!\pm} =  P \pm Q,
\label{109}
\end{equation}
where
\begin{eqnarray}
P\!\! &\doteq&\!\! \left(\frac{\varepsilon_2}{\varepsilon_c}EB + \frac{\varepsilon_3}{2\varepsilon_c}
E^2\cos\theta\right)\sin\theta,
\label{110}
\\
Q\!\! &\doteq&\!\! \frac{1}{\sqrt{\varepsilon_c\mu_c}}\left(\! 1\! -\!\frac{3}{2}\frac{\varepsilon_1}{\varepsilon_c}E^2\! 
-\! \frac{\varepsilon_2}{2\varepsilon_c}B^2\! -\! \frac{\varepsilon_3}{\varepsilon_c}EB\cos\theta\! \right).
\label{111}
\end{eqnarray}
Since $\varepsilon_i$ refers to small perturbations compared to the background permittivity $\varepsilon_c$,
we obtain $Q>0$ and $Q>P$. In this way, $v_e^{\!+}>0$ and $v_e^{\!-}<0$. 
Summarizing, we obtain that
$
\vec{v}_e^{\,+} = |v_e^{\!+}|\hat{z}
$ and 
$\vec{v}_e^{\,-} = -|v_e^{\!-}|\hat{z}$. 
In other words, when the $e$-ray propagates in the $\hat{z}$ direction it will present a phase velocity which is
generally different from the velocity of itself propagating in the opposite direction ($-\hat{z}$). 
Thus, for each direction $\hat{q}$ there will be only
one value for the velocity of the extraordinary ray, which propagates anisotropically.
Additionally, since the ordinary ray propagates isotropically, there will be 
birefringence phenomena in any direction but those where the velocity of both rays coincide.

With the usual definition of the refraction index ($n=1/v$) we obtain the following result for the 
maximum difference between the propagation of both rays  in the present situation
\begin{eqnarray}
n_{\|} - n_{\bot} &\simeq& \sqrt{\varepsilon_c\,\mu_c} \left[-\frac{\varepsilon_1}{\varepsilon_c}E^2 
+ \frac{\varepsilon_2}{\varepsilon_c}\sqrt{\varepsilon_c\,\mu_c}\, EB \sin\theta
\right.
\nonumber
\\
&& \left. - \frac{\varepsilon_3}{2\varepsilon_c}\left(EB - \sqrt{\varepsilon_c\,\mu_c} 
\, E^2\sin\theta\right)\cos\theta\right].  
\label{135}
\end{eqnarray}

It is interesting to note that, by  considering the wave vector in the 
opposite direction ($-\hat{q}$), the result presented in Eq. (\ref{135}) is modified as
\begin{eqnarray}
n_{\|} - n_{\bot} &\simeq& \sqrt{\varepsilon_c\,\mu_c} \left[-\frac{\varepsilon_1}{\varepsilon_c}E^2 
- \frac{\varepsilon_2}{\varepsilon_c}\sqrt{\varepsilon_c\,\mu_c}\, EB \sin\theta
\right.
\nonumber
\\
&&-\left. \frac{\varepsilon_3}{2\varepsilon_c}\left(EB + \sqrt{\varepsilon_c\,\mu_c} 
\, E^2\sin\theta\right)\cos\theta\right].  
\label{139}
\end{eqnarray}
In this case [Eq. (\ref{139})] the terms in $\varepsilon_1$ and $\varepsilon_2$ present the same sign. 
Thus, if $\varepsilon_1/\varepsilon_2>0$  and $\theta = \pi/2$ in the latter configuration of fields, 
the magnetoelectric contribution to the birefringence would magnify the effect. 
\\

\paragraph{Crossed fields:}
The case of crossed electric and magnetic fields ($\theta = \pi / 2$), which has been experimentally
measured, is obtained directly from Eq. (\ref{135}) as
\begin{eqnarray}
n_{\|} - n_{\bot} \simeq \sqrt{\varepsilon_c\,\mu_c} \left(-\frac{\varepsilon_1}{\varepsilon_c}E^2 
+ \frac{\varepsilon_2}{\varepsilon_c}\sqrt{\varepsilon_c\mu_c}\,EB\right).  
\label{140}
\end{eqnarray}
The term corresponding to the Cotton-Mouton effect does not appear 
since we have considered the particular case of non-magnetic medium ($\mu=\mu_c=$ constant). 
The inclusion of this term can be done simply by setting the permeability as a function of the magnetic field.

A kind of magnetoelectric birefringence will still occur when the medium behaves as $\varepsilon=\varepsilon(B)$ 
and both external fields $\vec{E}$ and $\vec{B}$ are present. In this case Eq. (\ref{140}) reduces to
$n_\| -n_\bot \simeq \mu_c\varepsilon_2 EB$, which shows that the 
magnetoelectric birefringence would be the dominant effect. In fact, it appears as the
unique effect.  This non-standard magnetoelectric birefringence occurs also in the more general
situation of non-crossed fields. In this case, the difference between the refraction indexes is
given by 
\begin{equation}
n_\| - n_\bot \simeq \mu_c\varepsilon_2 EB \sin\theta.
\end{equation}
We notice that its maximum value occurs when the fields are crossed ($\theta=\pi/2$).
\\

\paragraph{Jones birefringence:}
The Jones effect is another kind of magnetoelectric birefringence, also recently observed
\cite{roth2000}, which occurs with parallel electric $\vec{E}$ and magnetic $\vec{B}$ fields 
perpendicular to the wave vector $\vec{q}$. This case can be obtained directly from 
Eq. (\ref{135}) [or from Eq. (\ref{139})] and results in
\begin{eqnarray}
n_{\|} - n_{\bot} \simeq \sqrt{\varepsilon_c\,\mu_c} \left(-\frac{\varepsilon_1}{\varepsilon_c}E^2 
- \frac{\varepsilon_3}{2\varepsilon_c}EB\right).  
\label{141}
\end{eqnarray}

Magnetoelectric birefringence with crossed fields and Jones birefringence are 
very similar effects. Nevertheless, they are independent to each other. Note that the terms in 
the product $EB$ are proportional to different coefficients. In the case of crossed fields it
is proportional to the coefficient $\varepsilon_2$, which is coupled with the dependence on
$B^2$  in the permittivity. On the other hand, in the case of Jones birefringence, it is proportional
to the coefficient $\varepsilon_3$, coupled with the dependence on the product $EB$ in the
permittivity. 

\begin{figure}[!hbt]
\leavevmode
\centering
\includegraphics[scale = 0.7]{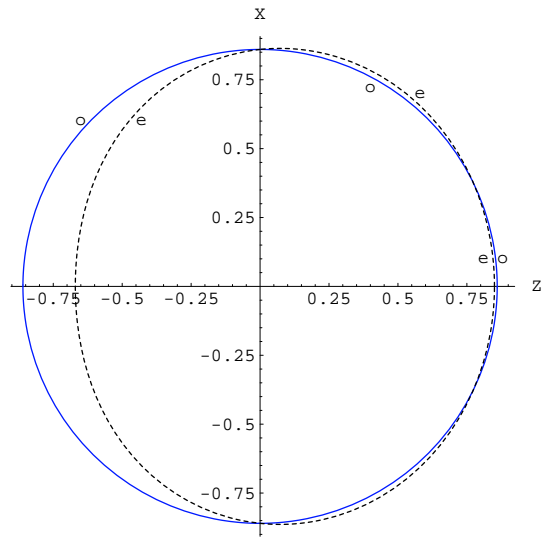}
\caption{{\small\sf Normal surfaces for the ordinary (circular solid line) and extraordinary (dashed line) 
rays propagating in an isotropic material media with dielectric coefficients 
given by $\varepsilon = \varepsilon(E,B)$ and  $\mu = \mu_c =$ constant. 
The symbols $e$ and $o$ stand for the front waves associated with
the $e$-ray and $o$-ray, respectively. The plot is based on Eqs. (\ref{131}) and (\ref{105}), where we have considered the
field configuration as $\hat{E}=\hat{x}$ and $\hat{B}=\hat{y}$, and also $\hat{q}\cdot\hat{B}=0$. The numerical
values were chosen as to satisfy the approximations assumed in Section \ref{ivb}.}}
\label{fig1}
\end{figure}

Another interesting aspect related to these effects is the form of the front wave at
a given instant of time. As we
have shown, for all cases the $e$-ray propagates anisotropically. Particularly, the 
magnitude of the effect is generally different in opposite directions. In order to visualize the
effect of this anisotropy in the propagation direction, let us analyze a particular case
where $(\hat{E}\hat{B})=0$, $(\hat{q}\hat{B})=0$ and $(\hat{q}\hat{E})=\cos\psi$. Let
us consider a Cartesian coordinate system such that $\hat{E}=\hat{x}$ and $\hat{B}=\hat{y}$, and
solve the equation for the curves representing the intersection of the normal surfaces with
the $XZ$-plane. Since the $o$-ray propagates isotropically, its associated normal 
surface is represented by a circle in this plane, at any instant of time.
On the other hand, the $e$-ray does not propagates isotropically and its normal surface will
be not symmetric with respect to the plane $XY$, as shown in Fig. \ref{fig1}.
We notice the possibility to the velocity of the $e$-ray be greater 
or smaller than the velocity of the $o$-ray, depending on the direction of propagation. Further,
depending on the parameters in the permittivity expansion, there will be more than
two directions in which the velocities of both rays coincide.
%
%

For any of the cases considered in this section the corresponding polarization vectors $e^\mu$ can be
directly obtained by returning each specific solution in the
set of equations (\ref{a1})-(\ref{a3}) and solving it for the coefficients $a$, $b$ and $c$.


\section{Conclusion}
\label{VI}
A tensorial formalism was here presented for the description of 
monochromatic electromagnetic waves inside
material media with nonlinear dielectric properties. The limit of geometrical optics was
considered. The eigen-vector problem for general media was presented and solved for
isotropic material media characterized by dielectric coefficients as
$\varepsilon_{\mu\nu}(E,B)$ and $\mu_{\mu\nu}(E,B)$. The solution for the wave propagation 
in isotropic media described by these coefficients was obtained directly from 
the general dispersion relation expressed by Eq. (\ref{58}). 
Using this main result, a detailed analysis of the birefringence phenomena was performed. 
Some well known effects were recovered and some new information was presented. 
The results include the description of the Kerr electro-optical effect, 
the Cotton-Mouton magneto-optical effect, and the recently measured magnetoelectric 
birefringences. 

It was shown that the Kerr birefringence in isotropic media is fundamentally 
related to the dependence of the permittivity on the electric field as $\varepsilon(E)$.
The application of an external magnetic field does not bring changes to the
effect. Symmetrically, the necessary condition for the appearance of the Cotton-Mouton effect 
is the dependence of the permeability on the magnetic field as $\mu(B)$. Similarly, an
applied electric field does not alter the results. 
 
In what refers to the magnetoelectric birefringence, some few remarks are in order.
First, from the general dispersion relation it was found three possible solutions for the
phase velocities. One is related to the propagation of the ordinary ray
and the other two are related to the propagation of the extraordinary ray. The two 
generally distinct values for the extraordinary ray are related to
the distinct values of its velocity in opposite directions, as it was explicitly shown for the
case where $\hat{q}$ is perpendicular to the plane that contains the 
electric and magnetic fields. For the case of parallel fields
it was found that the propagation is symmetric in opposite directions. In both cases the propagation
occurs anisotropically.  Another remarkable result is the birefringence effect occurring with
the permittivity $\varepsilon (B)$ and constant permeability $\mu_c$, in the presence of 
external fields. For this case, if the external electric field is absent there will be no 
birefringence. Nevertheless, if both the electric and the magnetic fields are present
a kind of magnetoelectric birefringence appears as the unique effect. This result
shows that it is possible to produce such birefringence effect without any
other accompanying standard birefringence, as the Kerr or Cotton-Mouton effects. By symmetry,
it can be inferred that a similar effect occurs with $\mu(E)$ and constant permittivity.
Finally, it should be stressed the behavior of the front wave associated with
the extraordinary ray. Depending on each medium in which the propagation occurs, 
the extraordinary ray can present its velocity greater or smaller than the velocity
of the ordinary ray depending on the direction of propagation. Further, the coincidence
between the both rays can occur in several directions. 

\acknowledgments
The authors are grateful to A. C. Zambroni de Souza for reading the 
manuscript. This work was partially supported by the Brazilian research agencies
CNPq and FAPEMIG. GPG thanks the support from FAPEMIG during his M-Sc 
studies.

\section*{Appendix}
In this appendix, the Kerr effect is derived by means of the Cayley-Hamilton method \cite{rodrigues}. 
Using the same assumptions presented in Section \ref{iva1} we obtain, from Eq. (\ref{23}),
\begin{eqnarray}
Z^\alpha{}_\tau = \left(\varepsilon - \frac{q^2}{\mu_c\omega^2} \right)h^\alpha{}_\tau - \varepsilon' E^\alpha E_\tau
-\frac{1}{\mu_c\omega^2}q^\alpha q_\tau.
\label{a.1}
\end{eqnarray}
The solutions of the eigenvalue problem stated by Eq. (\ref{5}) is formally given by 
$\det |Z^\alpha{}_\tau | =0$, which can be evaluated as
\begin{eqnarray}
(Z_1)^3- 3 Z_1 Z_2 + 2 Z_3 = 0,
\label{a.3}
\end{eqnarray}
with $Z_i$ ($i=1,2,3$) representing the traces:
\begin{widetext}
\begin{eqnarray}
Z_1 \doteq & Z^\alpha{}_\alpha =& 3\varepsilon - \frac{2q^2}{\mu_c\omega^2} +  \varepsilon' E^2,
\label{a.4}
\\
Z_2 \doteq & Z^\alpha{}_\tau Z^{\tau}{}_\alpha =& 3\left(\varepsilon - \frac{q^2}{\mu_c\omega^2} \right)^2
+2\left(\varepsilon' E^2 +\frac{q^2}{\mu_c\omega^2}\right)\left(\varepsilon - \frac{q^2}{\mu_c\omega^2} \right)
+\frac{2\varepsilon' (qE)^2}{\mu_c\omega^2} + \frac{q^4}{\mu_c^2\omega^4}+\varepsilon'^2 E^4,  
\label{a.5}
\\
Z_3 \doteq & Z^\alpha{}_\tau Z^{\tau}{}_\beta Z^\beta{}_\alpha =& 3\left(\varepsilon - \frac{q^2}{\mu_c\omega^2} \right)^3
+ 3\left(\varepsilon' E^2+\frac{q^2}{\mu_c\omega^2}\right)\left(\varepsilon - \frac{q^2}{\mu_c\omega^2} \right)^2 
+3\left[\varepsilon'^2 E^4 +\frac{2\varepsilon' (qE)^2}{\mu_c\omega^2}+\frac{q^4}{\mu_c^2\omega^4}\right]
\left(\varepsilon - \frac{q^2}{\mu_c\omega^2} \right)
\nonumber
\\ 
&&+ \varepsilon'^3 E^6 + \frac{3\varepsilon'^2 E^2 
(qE)^2}{\mu_c\omega^2} + \frac{3\varepsilon' q^2 (qE)^2}{\mu_c^2\omega^4} + \frac{q^6}{\mu_c^3\omega^6}.
\label{a.6}
\end{eqnarray}
\end{widetext}
Now, introducing the results (\ref{a.4})-(\ref{a.6}) in Eq. (\ref{a.3}), and using the definitions introduced
in Sections \ref{III} and \ref{IV}, we obtain
\begin{equation}
\left(1-\mu_c \varepsilon v^2\right)\left[\mu_c \varepsilon(\varepsilon + \varepsilon' E^2) v^2 
- \varepsilon -\varepsilon' E^2 (\hat{q}\cdot \hat{E})^2 \right]=0.
\label{a.8}
\end{equation}
As one can see, the above equation presents two solutions, which are the same solutions obtained 
in Section \ref{iva1} for the ordinary and extraordinary rays, as stated by Eqs. (\ref{71}) and (\ref{74}). 
Now, the Kerr birefringence is obtained in the same lines as discussed in the Section \ref{iva1}.


\begin{thebibliography}{88}
\bibitem{note1}
In the absence of sources the Maxwell equations can be represented by the
set of equations: $P^{\mu\nu}{}_{,\mu}=0$ and $\star{F}^{\mu\nu}{}_{,\mu}=0$,
where $P^{\mu\nu} = V^{\mu}D^{\nu} - V^{\nu}D^{\mu} + 
\eta^{\mu\nu\alpha\beta}V_{\alpha}H_{\beta}$
and $\star{F}^{\mu\nu} = V^{\mu}B^{\nu} - V^{\nu}B^{\mu} + 
\eta^{\mu\nu\alpha\beta}V_{\alpha}E_{\beta}$.
\end{thebibliography}
\end{document}